\documentclass[journal]{IEEEtran}

\usepackage{fst}
\usepackage{cite}

\usetikzlibrary{external}
\begin{document}

\title{Bit-Metric Decoding of Non-Binary LDPC Codes with Probabilistic Amplitude Shaping}
\author{Fabian Steiner,~\IEEEmembership{Student Member,} Georg B\"ocherer,~\IEEEmembership{Member,} and Gianluigi Liva,~\IEEEmembership{Senior Member}}

\tikzset{%
    funblock/.style = {draw,top color = gray!1,bottom color = gray!10,rounded corners,rectangle,text centered,minimum height = 20, minimum width = 25,inner sep=3}
}

\markboth{}{}%

\maketitle

\begin{abstract}
 A new approach for combining non-binary low-density parity-check (NB-LDPC)
 codes with higher-order modulation and probabilistic amplitude shaping (PAS) 
 is presented. Instead of symbol-metric decoding (SMD), a bit-metric decoder (BMD) is 
 used so that matching the field order of the non-binary 
 code to the constellation size is not needed, which increases
 the flexibility of the coding scheme. Information rates, density evolution thresholds and finite-length simulations
 show that the flexibility comes at no loss of performance if PAS is used.
\end{abstract}

\vspace{-.5\baselineskip}
\section{Introduction}
\label{sec:intro}

Higher-order modulation and advanced channel coding schemes play a central role for increasing the
\ac{SE} in next-generation communication systems. For instance, the upcoming
5G standard extended the range of modulation formats from 64-\ac{QAM} to 256-\ac{QAM}~\cite{3gpp-ts-38.211-v15.0.0}.
Non-binary codes are a natural candidate for \ac{FEC} schemes targeting higher-order modulation,
such as $M$-\ac{ASK} or $M$-\ac{QAM}, as the codeword symbols in the finite field $\setF_{2^p}$ can be
mapped directly to a sequence of constellation symbols, where $M = 2^m$ and $p = \ell\cdot m$, $\ell \in \setN$. The receiver uses \ac{SMD}~\cite{declercq2004regular,rong_combine_2008} for decoding.

\begin{figure}[t]
 \centering
 \footnotesize
 \includegraphics{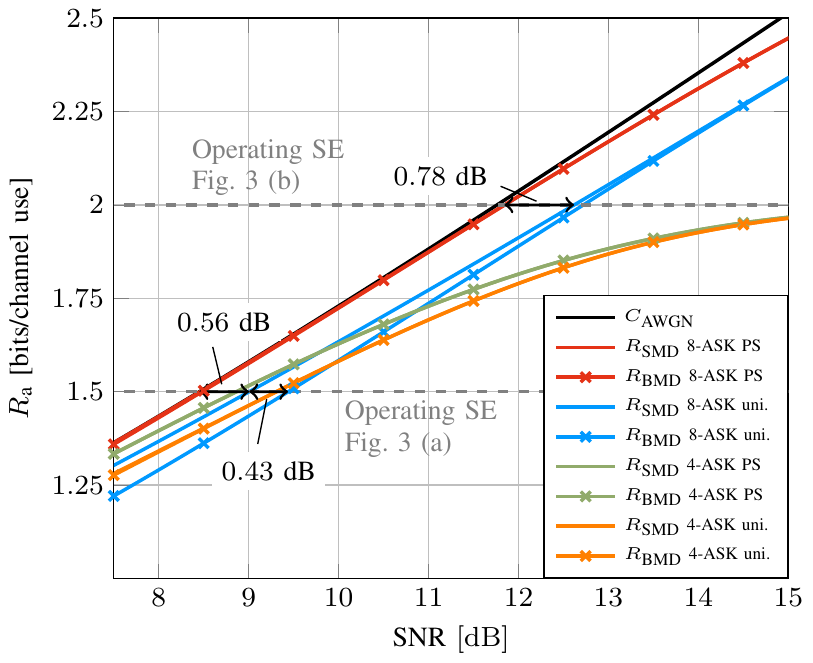}
 \caption{Information rates for \ac{BMD} and \ac{SMD} for both uniform and PS constellations.}
 \label{fig:rates8ask_smd_bmd_cmp}
\end{figure}

Most practical transponders use pragmatic schemes such as \ac{BICM}~\cite{guillen_i_fabregas_bit-interleaved_2008}
with binary codes and \ac{BMD}. \ac{BMD} ignores the correlation between the bit-levels forming one higher-order constellation symbol, 
which results in a loss of \SIrange{0.4}{0.5}{dB} for low to medium SNR ranges in the \ac{AWGN} channel (e.g., see the difference between the solid 
and dashed blue curves in Fig.~\ref{fig:rates8ask_smd_bmd_cmp}). As an important benefit, \ac{BMD} decouples the field size of the \ac{FEC} code
from the employed modulation order.

Probabilistic amplitude shaping (PAS)\acused{PAS} \cite{bocherer_bandwidth_2015} was introduced as a layered \ac{CM} architecture that combines 
\ac{PS} with binary \ac{FEC}. \ac{PAS} closes the gap to the Shannon limit and allows flexible rate
adaptation. Numerical results show that \ac{PAS} entails almost no loss even with \ac{BMD}.
Non-binary (NB)\acused{NB} codes show excellent performance for short blocklengths and low error rates~\cite{liva_code_2016}, which make them interesting candidates for ultra reliable
communication scenarios.  The authors of \cite{boutros_probabilistic_2017,steiner_ultra-sparse_2017} suggest extensions of \ac{PAS}
with \ac{NB} codes that enforce a relation between the modulation order and the field size of the \ac{NB} code. This property is not desired for flexible communication systems.

In this correspondance we show how \ac{NB-LDPC} codes can be operated with \ac{BMD} and \ac{PAS} to 
improve the flexibility for higher-order modulation and to avoid the \ac{BMD} loss. 
\ac{BMD} for \ac{NB-LDPC} codes was proposed in \cite{gorgoglione_binary_2012} to increase diversity in a fast Rayleigh fading scenario. In our work, the use of \ac{BMD} with \ac{NB-LDPC} codes is introduced in conjunction with \ac{PAS}, as a mean to achieve flexibility from two viewpoints. First, the de-coupling of constellation size and finite field order allows using codes constructed over large order finite fields with constellations of arbitrary cardinality without placing any constraints on the matching of the two. Second, the use of \ac{PAS} enables the achievement of large shaping gains and attaining a remarkable degree of flexibility with respect to transmission rates. This large flexibility comes at no visible performance loss with respect to \ac{SMD} applied to  \ac{NB-LDPC} codes \cite{boutros_probabilistic_2017,steiner_ultra-sparse_2017}. Our findings are validated by Monte Carlo \ac{DE}~\cite[Ch.~47.5]{mackay_IT} and finite length simulations with ultra-sparse \ac{NB-LDPC} codes.

\begin{figure*}
 \footnotesize
 \centering
 \includegraphics{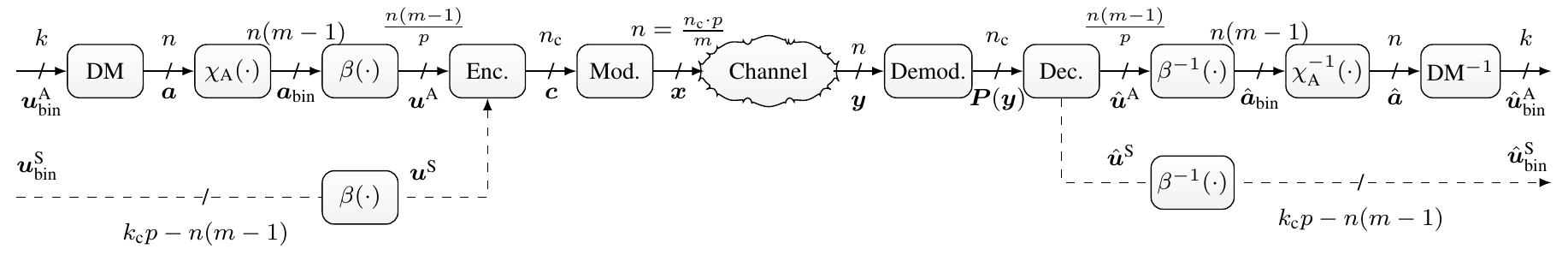}
 \caption{Operating a rate $R_\tc=k_\tc/n_\tc$ \ac{NB-LDPC} code with PAS and \ac{BMD}. The dashed lines are needed for code rates $R_\tc > (m-1)/m$~\cite[Sec.~IV-D]{bocherer_bandwidth_2015}.
 The functions $\chi_\tA(\cdot)$ and $\beta(\cdot)$ are applied to each amplitude in the vector $\va$ and chunks of $p$ consecutive bits in $\va_\tbin$, respectively. The decoder input
 is the matrix $\vP(\vy) = (\vP_1(\vy)^\tT, \vP_2(\vy)^\tT, \ldots, \vP_{n_\tc}(\vy)^\tT)$.}
 \label{fig:model_pas}
\end{figure*}

\vspace{-.6\baselineskip}
\section{Preliminaries}
\label{sec:prelim}

\subsection{System Model}
\label{sec:system_model}

Consider transmission over a real-valued \ac{AWGN} channel
\begin{equation}
 Y_i = X_i + Z_i\label{eq:system_model}
\end{equation}
for $i=1,\ldots,n$. The alphabet of the channel input $X_i$
is a scaled $M=2^m$-ary \ac{ASK} constellation $\cX = \{\pm1,\pm3,\ldots,\pm (M-1)\}$ such that $\E{X_i^2} = 1$. The results extend directly 
to \ac{QAM} by using \ac{ASK} for the in-phase and quadrature transmission. 
The noise $Z_i$ is a Gaussian random variable with zero mean and variance $\sigma^2$. The \ac{SNR} is $1/\sigma^2$.
As the channel is memoryless, we drop the index $i$ and denote the channel density as $p_{Y|X}$. 
The mutual information maximizing distribution under an average power constraint is a zero mean Gaussian input $X$ with unit variance,
and it yields the capacity expression
\begin{equation}
 \sfC_\tawgn(\text{SNR}) = \frac{1}{2}\log_2(1+\SNR)\label{eq:cap_awgn}.
\end{equation}
 In \cite{bocherer_achievable_2018}, it is shown that an achievable rate is
\begin{equation}
 R_\ta = \left[\entr(X) - \E{-\log_2\left(\frac{q(X,Y)}{\sum_{x\in\cX} q(x, Y)}\right)}\right]^+\label{eq:ra}
\end{equation}
where $\entr(X)$ is the entropy of the discrete \ac{RV} $X$, $[\cdot]^+ = \max(0,\cdot)$ and $q(x,y): \cX \times \setR \to \setR^+$ is the decoding metric.
For \ac{SMD}, the decoder uses the metric 
\begin{equation}
q(x,y) \propto P_{X|Y}(x|y)\label{eq:smd_metric}
\end{equation}
where $P_{X|Y}(x|y)$ is the conditional probability of the event $X = x$ when $Y=y$. The choice \eqref{eq:smd_metric} reduces \eqref{eq:ra} to the mutual information $\I(X;Y)$ between the channel input $X$ and channel output $Y$, i.e., we have
\begin{equation}
 R_\tsmd(\SNR; P_X) = \I(X;Y)\label{eq:rsmd}.
\end{equation}

For \ac{BMD}, we label each constellation point $x\in\cX$ with an $m$-bit binary label, 
i.e., $\chi : \cX \to \{0,1\}^m$ and $\chi(x) = b_1b_2\ldots b_m = \vb$. Its inverse is
$\chi^{-1}: \{0,1\}^m \to \cX$. A \ac{BRGC}~\cite{gray1953pulse} usually 
performs well for \ac{BMD} and the \ac{BMD} decoder uses the metric
\begin{equation}
 q(x,y) = \tilde q(\vb,y) \propto \prod_{i=1}^m P_{B_i|Y}(b_i|y)\label{eq:bmd_metric}.
\end{equation}
The choice~\eqref{eq:bmd_metric} reduces \eqref{eq:ra} to
\begin{equation}
 R_\tbmd(\SNR; P_X) = \left[\entr(\vB) - \sum_{i=1}^m \entr(B_i|Y)\right]^+\label{eq:rbmd}.
\end{equation}

\subsection{Non-Binary LDPC Codes}
\label{sec:nb_ldpc_intro}
A \ac{NB-LDPC} code $\cC$ is defined as the nullspace of the sparse parity-check matrix $\vH$ of dimension $m_\tc\times n_\tc$ 
where the non-zero entries of $\vH$ are taken from a finite field $\setF_{q}$, i.e., 
$\cC = \left\{\vc\in\setF_{q}^{n_\tc}: \vc\vH^{\tT} = \zeros\right\}$. In the following, we consider only extension fields of the Galois field $\setF_2$, i.e.,
we consider $\setF_q$ where $q = 2^p$. The primitive element of $\F_q$ is referred to as $\alpha$. The number of non-zero elements in each column $i \in \{1,\ldots,n_\tc\}$ (row $j\in \{1,\ldots,m_\tc\}$) is
refered to as the corresponding variable node degree $d_{\tv,i}$ (check node degree $d_{\tc,j}$).
In the following, we use a special class of \ac{NB-LDPC} codes, namely ultra-sparse regular LDPC codes~\cite{poulliat_design_2008}, which
have a constant variable node degree of $d_{\tv,i} = d_\tv = 2$ and a constant check node degree $d_\tc$. Their design rate $R_\tc$
is therefore $1-2/d_\tc$. We consider a full rank $\vH$ in the following and perform probability-domain based decoding~\cite{declercq_decoding_2007}.
Decoding approaches for NB-LDPC codes with lower complexity are discussed in, e.g., \cite{voicila_low-complexity_2010}. 
They are also applicable for the proposed BMD. We also introduce the mapping
$\beta(\cdot)$ which relates a length $p$ binary string to a field element in $\setF_q$, i.e.,
\begin{equation}
 \beta: \{0,1\}^p \to \setF_q\label{eq:beta}.
\end{equation}
Its inverse $\beta^{-1}(c)$ for $c\in\setF_q$ is the binary image of $c$.

\subsection{Probabilistic Amplitude Shaping (PAS)}
\label{sec:pas_intro}
\ac{PAS} is a \ac{CM} scheme that combines \ac{PS} with \ac{FEC}~\cite{bocherer_bandwidth_2015}. It builds upon
two important properties. First, the capacity achieving distribution $P_X^*$ for the \ac{AWGN} channel is symmetric. We therefore
factor the input distribution into an amplitude and sign part as $P_X(x) = P_A(\abs{x}) \cdot P_S(\sign(x))$, where $P_A$ is non-uniform
on $\cA = \{\abs{x}, x\in\cX\}$ and $S$ is uniform on $\{-1,+1\}$. Second, the scheme exploits systematic encoding to preserve the non-uniform $P_A$.
It copies the amplitudes (or a representation thereof) into the information part of the codeword and uses the approximately uniform distributed parity bits as signs. As a result, PAS requires \ac{FEC} code rates with $R_\tc \geq (m-1)/m$~\cite[Sec.~IV-B, IV-D]{bocherer_bandwidth_2015}.

In the following, we distinguish between sign and amplitude bit labels. For this, we introduce an amplitude labeling function
$\chi_\tA: \cA \to \{0,1\}^{m-1}$ such that $\chi(x) = (b_1, b_2, \ldots, b_m) = (b_1, \chi_\tA(\abs{x}))$, i.e., the sign bit is
placed in the first bit-level.

The \ac{DM}~\cite{schulte_constant_2016} realizes the non-uniform distribution $P_A$ on the amplitude symbols. It takes $k$ uniformly distributed
input bits and maps them to a length $n$ sequence of symbols with a specified empirical distribution. For PAS, the output set is the set of amplitude values $\cA = \{1, 3, \ldots, M-1\}$.
The \ac{DM} rate is $R_\tdm = k/n$. The transmission rate is~\cite[Sec.~IV-D]{bocherer_bandwidth_2015}
\begin{equation}
 \eta = R_\tdm + 1 - (1-R_\tc)\cdot m\label{eq:Rtx_PAS}.
\end{equation}

\section{Symbol-Metric Decoding of NB-LDPC Codes}
\label{sec:smd}

\subsection{SMD for Uniform Signaling}
For a given $M=2^m$-ASK signalling constellation, we choose a $\setF_q$ code with
$q = \ell\cdot m, \ell \in \setN$, such that a length $\ell$ sequence of constellations
points can be mapped exactly to one $\setF_q$ symbol using a mapping function
$\beta_{\cX}: \cX^\ell \to \setF_q$. The soft information for the decoder is given by a length-$q$ vector at the $i$-th variable node
by \begin{equation}
 \vP_i(\vy) = \left(P_i(\vy,0), P_i(\vy,1), \ldots, P_i(\vy,\alpha^{q-2})\right).\label{eq:dec_pi}
\end{equation}
where $\vy = (y_1, \ldots, y_\ell)$. The vector entries 
$P_i(\vy, c)$ denote the probability that the $i$-th codeword symbol is $c$ given that the associated
receive sequence is $\vy$. It is calculated as
\begin{equation}
 P_i(\vy,c) \propto \prod_{j=1}^\ell p_{Y|X}(y_j|[\beta_{\cX}^{-1}(c)]_j)\label{eq:pi_smd_uni}.
\end{equation}

\vspace{-2\baselineskip}
\subsection{SMD for PAS}
\label{sec:smd_pas}
For \ac{PAS}, a scheme with \ac{NB-LDPC} codes and \ac{SMD} was introduced in \cite{steiner_ultra-sparse_2017}, which ensures
that the desired amplitude distribution is not changed after encoding. This property 
can be achieved by mapping a length $\ell \in \setN$ sequence of amplitudes (each amplitude is represented by $(m-1)$ bits)  to 
one $\setF_q$ symbol

and encoding them systematically. The bits forming the parity symbols (as well as potentially additional ones from the information 
part for $R_\tc > (m-1)/m$) are used as signs for the amplitudes to form the channel inputs.

The soft information for the \ac{NB-LDPC} decoder with \ac{SMD}
is calculated as shown in \cite[Eqs. (9) and (10)]{steiner_ultra-sparse_2017}.
This approach enforces the condition $p = \ell \cdot (m-1)$ between
the \ac{NB} code and the underlying constellation size.

\section{Bit-Metric Decoding of NB-LDPC Codes}
\label{sec:bmd}

\subsection{BMD for Uniform Constellations}
\label{sec:bmd_uni}

We now describe how a \ac{NB-LDPC} code can be operated with \ac{BMD}.
The blockwise application of \eqref{eq:beta}
maps a length $k_\tc \cdot p$ vector of uniformly distributed bits
to $k_\tc$ symbols of $\setF_q$. This sequence is
encoded into a length $n_\tc$ symbols codeword $\vc$ with binary representation
$\vc_\tbin$.
Eventually, the modulation mapper maps blocks of $m$ bits to one $2^m$-ASK symbol
\[
 x_i = \chi^{-1}(c_{\text{bin},(i-1)\cdot m + 1},\ldots,c_{\text{bin},i\cdot m}), \quad i = 1,\ldots,n.
\]

At the receiver side, the received sequence is demodulated by calculating the
entries $l_{i,j}$ of the soft information vector $\vl$
\begin{align}
 l_{i,j} &=  \log\left(\frac{P_{B_j|Y}(0|y_i)}{P_{B_j|Y}(1|y_i)}\right)\label{eq:llrs} 
\end{align}
for $i = 1,\ldots, n$ and $j = 1,\ldots, m$. The distribution $P_{B_j|Y}(b|y)$
is 
\begin{equation*}
 P_{B_j|Y}(b|y) \propto \sum_{x \in \cX_j^b} p_{Y|X}(y|x) P_X(x)
\end{equation*}
where $\cX_j^b = \{x \in \cX: [\chi(x)]_j = b\}$. The input \eqref{eq:dec_pi} 
to the \ac{NB-LDPC} decoder is calculated as
\begin{equation}
P_i(c) = \frac{\tilde P_i(c) }{\sum_{c'\in\setF_q} \tilde P_i(c')}\quad \text{with} \quad \tilde P_i(c) = \prod_{j=1}^p \tilde P_{ij}\label{eq:p_decode1}
\end{equation}
for $i = 1, \ldots, n_\tc$ and $j = 1, \ldots, p$, where
\begin{align}
\tilde P_{i,j} = \begin{cases}
       \frac{\exp(l_{i,j})}{1+\exp(l_{i,j})}, & \text{if } [\beta_{\setF_q}^{-1}(c)]_j = 0,\\
       \frac{1}{1+\exp(l_{i,j})}, & \text{if } [\beta_{\setF_q}^{-1}(c)]_j = 1.
 \end{cases}\label{eq:p_decode2}
\end{align}

\vspace{-\baselineskip}
\subsection{BMD for PAS}
\label{sec:bmd_pas}

\begin{figure*}
\centering
 \footnotesize
 \subfloat[][SE = \SI{1.5}{\bpcu}, $n_{\tc,\tbin} = 576$]{\includegraphics{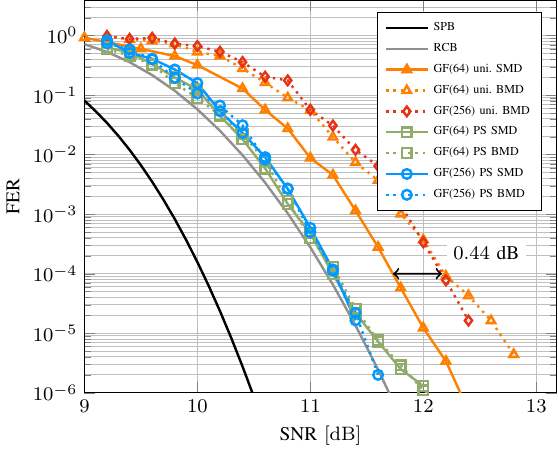}}
 \subfloat[][SE = \SI{2.0}{\bpcu}, $n_{\tc,\tbin} = 576$]{\includegraphics{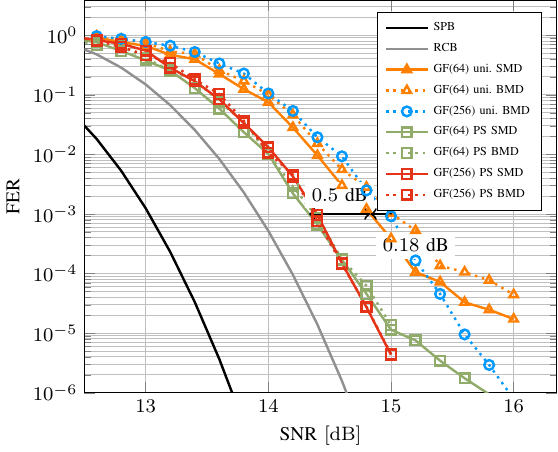}}
 \subfloat[][SE = \SI{3.0}{\bpcu}, $n_{\tc,\tbin} = 1152$]{\includegraphics{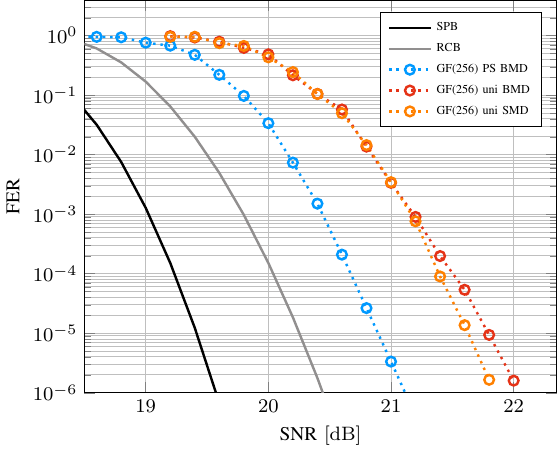}}
 \caption{Performance comparison of NB-LDPC codes for different \acp{SE} and decoding metrics.}
 \label{fig:plots}
\end{figure*}

The same principle as shown in Sec.~\ref{sec:bmd_uni} can also be applied to \ac{PAS}
and is shown in Fig.~\ref{fig:model_pas}. A number of $k$
uniformly distributed information bits are matched to $n$ amplitudes following a 
specified distribution. Using the amplitude mapping $\chi_\tA$ the amplitudes
are mapped to a length $n\cdot(m-1)$ bit string, mapped to $\setF_q$ symbols and encoded into 
the codeword $\vc$. A modulator then maps the binary image of $\vc$ to channel inputs $x \in \cX$ via
a consecutive application of $\chi^{-1}$.

At the receiver side, the demapper calculates a soft information vector as shown in
\eqref{eq:llrs}, \eqref{eq:p_decode1} and \eqref{eq:p_decode2} for the uniform scenario.

\emph{Example.}
 Consider a length $n_\tc = 3$, rate $R_\tc = 2/3$ code over $\setF_{32}$ ($p = 5$), while
 using an 8-ASK constellation ($m = 3$) such that the channel is used $n = (n_\tc\cdot p)/m = 5$ times
 with constellation symbols $x_1, x_2, x_3, x_4, x_5$. The length $m$ binary label of the $i$-th channel
 symbol is referred to as $b_{i,1}\ldots b_{i,m}$. That is, for the given scenario, we have
 $\chi(x_i) = b_{i,1}b_{i,2}b_{i,3}$.
 Conventional \ac{PAS} with \ac{NB} codes and \ac{SMD}~\cite{steiner_ultra-sparse_2017} is not possible for these parameters, as $p = 5$ is not an integer multiple of $m-1 = 2$.
 After encoding, the binary image of the codeword $\vc = (c_1, c_2, c_3)$ is
 \[
 \small
  \vc_{\tbin} = (\underbrace{b_{1,2}b_{1,3}b_{2,2}b_{2,3}b_{3,2}}_{\beta_{\setF_{32}}(c_1)}, \underbrace{b_{3,3}b_{4,2}b_{4,3}b_{5,2}b_{5,3}}_{\beta_{\setF_{32}}(c_2)}, \underbrace{b_{1,1}b_{2,1}b_{3,1}b_{4,1}b_{5,1}}_{\beta_{\setF_{32}}(c_3)}).
 \]
 The binary image of the parity symbol $c_3\in\setF_{32}$ provide the signs for
 the five channel uses and the soft information vector reads as
 \begin{multline*}
  \vl = \left(l_{1,2}l_{1,3}l_{2,2}l_{2,3}l_{3,2} l_{3,3}l_{4,2}l_{4,3} l_{5,2} l_{5,3} l_{1,1}l_{2,1}l_{3,1}l_{4,1}l_{5,1}\right).
 \end{multline*}

 Eventually, the vector $\vl$ is combined as shown in \eqref{eq:p_decode1} and \eqref{eq:p_decode2}
 to form the decoder a-priori soft-information.

\section{Simulation Results}
\label{sec:results}

We now present numerical simulation results that target \acp{SE} of \SI{1.5}{bits per channel use (\bpcu)}, \SI{2.0}{\bpcu} for 8-ASK and \SI{3.0}{\bpcu} for 16-ASK, respectively.
As benchmark curves, we plot Shannon's sphere packing (SP) bound~\cite{shannon_probability_1959} and Gallager's
Random Coding bound (RCB)~\cite[Theorem~5.6.2]{gallager1968}. We evaluate the later for the shaped distributions which are also used by the 
demapper. The considered modes and the employed FEC code rates are summarized in the first rows of Table~\ref{tab:de_results}.

The transmission rate is $\eta = R_\tc m$ for uniform signaling and \eqref{eq:Rtx_PAS} for PAS.
For a given code rate, we can adjust the matcher rate $R_\tdm$ to achieve a desired transmission rate. We use \ac{MB} distributions~\cite{kschischang_pasupathy_maxwell}
of the form $P_X(x) \propto \exp(-\nu x^2)$. Numerical results indicate that \ac{MB} distributions also perform well for BMD, see, e.g., \cite[Table~3]{bocherer_bandwidth_2015}.

All codes are ultra-sparse \ac{NB-LDPC} codes over $\setF_{64}$ or $\setF_{256}$ with a regular variable node degree of $d_\tv = 2$. The non-zero entries of the $\setF_{64}$ codes have been optimized row-wise via the binary-image
method of \cite{poulliat_design_2008}, while the entries of the $\setF_{256}$ have been chosen randomly. The error floor for some $\setF_{64}$ codes is caused by low weight codewords and can be mitigated by
ensuring the full rank condition of \cite{poulliat_design_2008}.  
Observe that the finite length \ac{SMD} and \ac{BMD} \ac{FER} performance (after the inverse DM) of the \ac{PAS} schemes coincide in Fig.~\ref{fig:plots} (a) and (b) for all considered codes. 
This is also reflected in the Monte Carlo \ac{DE} thresholds~\cite[Ch.~47.5]{mackay_IT} of Table~\ref{tab:de_results}. Fig.~\ref{fig:plots}~(c) shows a setting where \ac{BMD} improves the flexibility of the modulation setup, as 
\ac{PAS} with \ac{SMD} can not be operated with 16-ASK and a \ac{NB} code over $\setF_{256}$. Using \ac{BMD} circumvents this restriction.
As expected from the \ac{DE} thresholds, the performance of \ac{BMD} for the uniform cases is degraded compared to \ac{SMD} (compare Fig.~\ref{fig:plots} (a), (b) 
orange solid and dotted) for low code rates, but become similar for higher ones. In particular, for an information rate of \SI{1.5}{bpcu} in Fig.~\ref{fig:rates8ask_smd_bmd_cmp}, uniform 8-ASK has a higher BMD loss than shaped 8-ASK.
For information rates above \SI{1.5}{bpcu} the BMD loss of uniform 8-ASK decreases, while the gap to capacity becomes larger.
In the PAS implementation, shaped 8-ASK uses a higher linear FEC code rate than uniform 8-ASK for the same information rate.  In this sense, PAS allows to combine a small gap to capacity with a low BMD loss by using high rate FEC code rates.


\begin{table}[t]
\centering
\caption{Density Evolution Thresholds and required asymptotic SNR values for (5) and (7) in \textnormal{\si{dB}}}
\label{tab:de_results}
\begin{tabular}{lllllll}
  \toprule
      & \multicolumn{4}{c}{8-ASK} & \multicolumn{2}{c}{16-ASK}\\
      & \multicolumn{2}{c}{SE = \SI{1.5}{\bpcu}} & \multicolumn{2}{c}{SE = \SI{2}{\bpcu}} & \multicolumn{2}{c}{SE = \SI{3}{\bpcu}}\\
  \midrule
    $R_\tc$  & 1/2 & 3/4 & 2/3 & 3/4 & 3/4 & 5/6\\
    Mode & uni. & PAS & uni. & PAS & uni. & PAS\\
  \midrule
  $R_\tbmd^{-1}$  [\si{dB}] & 9.44 & 8.48  & 12.72 & 11.89 & 19.25 & 18.11\\
  $R_\tsmd^{-1}$  [\si{dB}] & 9.00 & 8.46  & 12.61 & 11.87 & 19.17 & 18.10\\
  \midrule
  $\setF_{64}$, BMD [\si{dB}] &  9.93                   & 8.90 & 13.20                                        & 12.31 & -- & --\\
  $\setF_{64}$, SMD [\si{dB}] &  9.53                   & 8.92 & 13.10                                        & 12.29 & -- & --\\
  \midrule
  $\setF_{256}$, BMD [\si{dB}] & 9.91                    &  8.93 & 13.20                                      & 12.31 & 19.79 & 18.54\\
  $\setF_{256}$, SMD [\si{dB}] & --                    &  8.93 & --                                           & 12.31 & 19.85 & --\\
  \bottomrule
 \end{tabular}
\end{table}

\vspace{-1\baselineskip}
\section{Conclusion}
\label{sec:conclusion}

We have shown that \ac{BMD} of \ac{NB-LDPC} codes with
\ac{PAS} achieves the same performance as \ac{SMD}.
Numerical simulation results confirm the information
rate and \ac{DE} threshold analysis.
\ac{BMD} of \ac{NB-LDPC} codes with PAS increases the flexibility in code design as any field order can be combined
with any modulation size. This is particularly important if \ac{NB} codes over smaller field orders are designed for \ac{PAS}, e.g., for $\setF_{32}$, which could only be operated with 64-ASK in case of \ac{SMD}. This would decrease the
decoding complexity significantly, while a careful code design is expected to provide similar performance as codes over high order fields. 

\vspace{-1.0\baselineskip}

\end{document}